\documentclass[preprint,review,1p,times]{elsarticle}

\usepackage[latin1]{inputenc}
\usepackage{graphics}
\usepackage{epsfig}
\usepackage{epstopdf}
\usepackage[active]{srcltx}
\usepackage[all]{xy}
\usepackage[ruled,vlined]{algorithm2e}
\usepackage{lineno}
\usepackage{hyperref}
\usepackage{amsmath,amssymb,amsfonts,amstext}

\journal{arXiv}

\def \aa {\mathbf{a}}
\def \bb {\mathbf{b}}

\def \R {\mathbb{R}}
\def \Z {\mathbb{Z}}

\newcommand{\C}{\mathcal{C}}
\newcommand{\G}{\mathcal{G}}
\newcommand{\OO}{\mathcal{O}}

\newtheorem{teo}{Theorem}[section]

\newcommand{\dem}{\noindent \textsl{Proof}: $\,$}
\newcommand{\fim}{\hfill $\rule{2.0mm}{2.0mm}$ \\}    

\newcommand{\vetn}[2]{\mathbf{#1}_1,\dots ,\mathbf{#1}_#2}
\newcommand{\vv}[1]{\mathbf{#1}}

\begin{document}

\begin{frontmatter}

\title{Optimum Commutative Group Codes}

\author[fca]{Cristiano Torezzan\fnref{cris}}
\address[fca]{School of Applied Sciences, University of Campinas, SP, Brazil}
\ead{cristiano.torezzan@fca.unicamp.br}
\fntext[cris]{FAPESP: 05/58102-7}
\cortext[cor1]{Correspondence author}
\author[fca]{João E. Strapasson\fnref{jes}}
\ead{joao.strapasson@fca.unicamp.br}
\fntext[jes]{FAPESP: 07/00514-3}
\author[imecc]{Sueli I. R. Costa\fnref{sueli}\corref{cor1}}
\address[imecc]{Institute of Mathematics, University of Campinas, SP, Brazil, 13.083-859}
\ead{sueli@ime.unicamp.br}
\fntext[sueli]{FAPESP: 02/07473-7, CNPq: 304573/2002}
\author[roger]{Rogerio M. Siqueira}
\address[roger]{School of Arts, Science and Humanities, University of São Paulo, Sao Paulo, Brazil}
\ead{rogerms@usp.br}



\begin{abstract}
A method for finding an optimum $n$-dimensional commutative group code of a given order $M$ is presented. The approach explores the structure of lattices related to these codes and provides a significant reduction in the number of non-isometric cases to be analyzed. The classical factorization of matrices into Hermite and Smith normal forms and also basis reduction of lattices are used to characterize isometric commutative group codes. Several examples of optimum commutative group codes are also presented.
\end{abstract}


\end{frontmatter}


\section{Introduction}

The design of spherical codes for signal transmission through a Gaussian channel is a classical problem in coding theory, where group codes have proved useful \cite{Ing03, Lo} since their appearance in the pioneering work of Slepian \cite{Sle}. The special attention devoted to these codes is largely due to their symmetry and homogeneity which arise from their special algebraic structure \cite{Sle2, forney}. The interest in such group codes has persisted with various studies have been developed \cite{Bene, caire, fag, fag2, for2, for3}, including some proposing applications in turbo concatenated and low density schemes \cite{Garello, Garin, Como08, Erez, Sridhara}. Recently it has been shown that the Shannon capacity of certain important channels, as the AWGN channel with $m$-PSK modulation, can be achieved using \textit{commutative group codes} \cite{Como09} and they will be focused here.

One of the underlying difficulties in the design of a group code is the finding of an initial vector which maximizes the minimum distance of the associated code, for a fixed group of orthogonal matrices; the so called \textit{initial vector problem}. This problem still does not have a general solution, although various important cases have been studied, including reflexion group codes \cite{mitt} and permutation group codes \cite{zino}. Besides, Biglieri and Elia have shown \cite{BigEl03} that for cyclic group codes the problem can be formulated as a linear programming problem. Here we extend their ideas and show that for any commutative group code, the initial vector problem can also be solved in the same way.

Furthermore, this paper deals with the more general problem of determining an optimum commutative group code in $\R^n$ for a given order $M$. We derive a two-step algorithm which leads to the finding of  a code with maximum minimum distance for a fixed number of points $M$. Our approach explores the connection between even dimensional commutative group codes and lattices related to them in the half of the dimension \cite{SIQ, CIRC}. Using basis reduction of lattices and the classical factorizations of matrices into Hermite and Smith normal forms, we characterize a set of relevant cases to be analyzed, after discarding isometric codes. The reduction process presented here can also be used in the solution of other problems where lattices \cite{everywhere}, in particular orthogonal sub-lattices, are involved; including coding and decoding process \cite{banihas,banihas2,Blake96, viterbi}, image compression \cite{Neel}, spherical codes on torus layers \cite{isit2009} and also the enticing lattice based cryptography \cite{mic1, bers, buc}.

This paper is organized as follows. Commutative group codes and some of their properties are presented in Section $2$. We then discuss the initial vector problem for those codes and characterize it as a linear programming problem in Section $3$. The main results are presented in Section $4$, where we prove a simple, but useful, extended Hermite normal form (theorem \ref{FNH}) which allows the characterization of isometric lattices by coordinate permutation; in this section we also derive theorems \ref{Class} and \ref{gera} which provide a significant reduction in the number of codes to be checked in the search for an optimum one. Our method is presented as a pseudo-code (\textbf{Algorithm 1}), and some examples of optimal codes in several dimensions are given.

\section{Commutative group codes}

Let $\OO_n$ be the multiplicative group of or\-tho\-go\-nal matrices $n \times n$ and $\G_n(M)$ be the set of all order $M$ commutative subgroups in $\OO_n$.

A \textit{commutative group code} $\C$ is a set of $M$ vectors which is the orbit of an initial vector $x_0$ on the unit sphere $S^{n-1} \subset \R^n$ by a given $G \in \G_n(M)$, i.e. 

$$\C := Gx_0 = \left\{ g x_0, g \in G \right\}.$$

We assume that $\C$ is substantial, i.e., not contained in a hyperplane.

The \textit{minimum distance} in $\C$ is defined as:

$$
\displaystyle d  := \min_{
\scriptsize
\begin{array}{c}
x,y \in \C \\ 
x \neq y
\end{array} } ||x-y|| = \min_{\scriptsize
\begin{array}{c}
g_i \in G \\ 
g_i \neq I_n
\end{array}} ||g_i x - x||,
$$
where $\displaystyle ||.||$ and $I_n$ denote the standard Euclidean norm and the identity matrix of order $n$ respectively.

In what follows, $\C(M,n,d)$ denotes a code $\C$ in $\R^n$ with $M$ points and minimum distance equal to $d$. A $\C(M,n,d)$ is said to be \textit{optimum} if $d$ is the largest minimum distance for a fixed $M$ and $n$.

As is well known, the minimum distance of a group code $\C$, generated by a finite group $G$, may vary significantly depending on the choice of the initial vector $x_0$. Therefore, the search for an optimum $n$-dimensional commutative group code with $M$ points requires the consideration of all $G \in \G_n(M)$ and solution of the initial vector problem for each $G$.

A well known real-irreducible representation of a finite commutative group of orthogonal matrices $G$ can be stated as follows:

\begin{teo}(\cite{gan} Theorem 12.1) Every commutative group $G \in \G_n(M)$ can be carried by the same real orthogonal transformation $q$ into a pseudo-diagonal form:
\label{teogan}
{

$$
	qg_{_i}q^{t}=[R_1(i),\ldots,R_k(i),\mu(i)_{2k+1},\ldots,\mu(i)_{n}]_{n\times n},
$$

\begin{equation}
\label{pseudo_diagonal}
\mbox{where  }R_{j}(i)=\left[
\begin{array}
[c]{cc}%
\cos(\frac{2\pi b_{ij}}{M}) & -\sin(\frac{2\pi b_{ij}}{M})\\
\sin(\frac{2\pi b_{ij}}{M}) & \cos(\frac{2\pi b_{ij}}{M})
\end{array}
\right],
\end{equation}

$$b_{ij} \in Z, \ \ 0 \leqslant b_{ij} \leqslant M \mbox{ and }\mu(i)_{l}=\pm 1  \mbox{, } l = 2k+1, \hdots, n, \, j = 1, \hdots, k, \forall g_i \in G.$$
}
\end{teo}

\section{The initial vector problem}
\label{IVP}
In this section we consider, for each group $G \in \G_n(M)$, the search for a vector $x$ in $S^{n-1}$ which maximizes the minimum distance between two points in $\C = Gx $, i.e., the search for an $x$ that solves:

$$
{\max_{x\in S^{n-1}}} \left( {\min_{g_i \in G, g_i \neq I_n}} ||g_i x - x||^2 \right)
$$
This initial vector problem has been solved only in certain special cases. Biglieri and Elia have shown in \cite{BigEl03} that, for cyclic groups, this search can be reduced to a linear programming problem (LP). Here, we extend their ideas and present an alternative formulation which also allows the reduction of the initial vector problem to a LP for any commutative group code.

According to Theorem \ref{teogan}, we have:

$$
||g_i x - x||^2 = 2 - 2 \left( \sum_{j=1}^{k} \left( 1 - 2 \sin^2(\frac{\pi}{M}b_{ij}) \right)(x_{2j-1}^2+x_{2j}^2) + \sum_{j=k+1}^{n-k}  \mu(i)_j (x_j^2) \right). 
$$
Considering 

$$y_j= \left\{ \begin{array}{cl} x_{2j-1}^2+x_{2j}^2 & \mbox{, if } j=1,\ldots,k \\ 
x_{j+k}^2 & \mbox{, if }j=k+1,\ldots,n-k \end{array}\right. ,$$ 
we obtain 

$$
||g_i x - x||^2 = 2 - 2 \left( \sum_{j=1}^{k} \left( 1 - 2 \sin^2(\frac{\pi}{M}b_{ij}) \right) y_j + \sum_{j=k+1}^{n-k}  \mu(i)_j y_j \right). 
$$
Thus, ${\displaystyle \max_{x\in S^{n-1}}} \left( \displaystyle {\min_{g_i \neq I_n}} ||g_i x - x||^2 \right)$ is equivalent to

$$ \max \min \left(  2 - 2 \left( \sum_{j=1}^{k} \left( 1 - 2 \sin^2(\frac{\pi}{M}b_{ij}) \right) y_j + \sum_{j=k+1}^{n-k}  \mu(i)_j y_j \right) \right),$$

$$\mbox{ subject to } \sum_{j=1}^{n-k} y_j =1, y_j \geq 0.$$
This \textit{max min} problem, which is linear in $y$, can be reduced to the following linear programming problem:  

$$
\max{z},
$$
subject to 

\begin{equation*}
\left\{
\begin{split}
 z  \leq &   2 - 2 \left( \sum_{j=1}^{k} \left( 1 - 2 \sin^2(\frac{\pi}{M}b_{ij}) \right) y_j + \sum_{j=k+1}^{n-k}  \mu(i)_j y_j \right)\\
 \displaystyle \sum_{k=1}^{n-k}y_i  = & 1 \\
 y_i \geq & 0 \\
\end{split}
\right.
\end{equation*}

Therefore, the initial vector problem for commutative group codes is equivalent to a linear programming problem with $n-k+1$ variables. Due to the symmetry of the function $\sin^2(x)$, in the case where the group $G$ is free of $2 \times 2$ reflection blocks ($n=2k$), the number of constraints can be reduced to $\left(\left\lfloor \frac{M}{2} \right\rfloor+1  \right)$.

\section{Optimum commutative group codes}

In this section we consider a more general problem of finding a commutative group code of order $M$ in $\R^n$ which has the largest minimum distance. To do this, we must consider all commutative groups $G \in \G_n(M)$ with the respective best initial vectors and compare the minimum distances of the correspondent codes.

Let us start by estimating the number of commutative group codes to be checked in order to find an optimum one.

As usual, we say that two groups $G$ and $H$ are equivalent if they are conjugate, i.e., 

$$G \approx H \Longleftrightarrow \exists \ \ p \in\OO_n; H=p\ G\, p^t.$$

Although the set $\G_n(M)$ is infinite, conjugate groups generate isometric codes. Specifically, given an initial vector $x$, $G \in \G_n(M)$ and $p \in \OO_n$, the group code generated by $G$ is isometric to the group code generated by $H = p\, G\, p^t$, with initial vector $p\,x$. In fact, for each $h_i = p g_i p^t \in H$, we must have

$$\| h_i(p\, x)-p\,x\| = \| (p\,g_i\,p^t)\,(p\, x)-p\,x\| =\| p\,g_i \, x-p\,x\| =\| g_i\, x-x\|.$$

Thus, the search for optimal commutative group codes can be restricted to groups which are distinct up to conjugacy. In other words, it is sufficient to consider just one representative for each class of the quotient $\G_n(M) / \approx$, resulting in a finite set. In fact, by Theorem \ref{teogan}, for each $G \in \G_n(M)$ there exists $H = qGq^t$ in a pseudo-diagonal form i.e., for each class in quotient $\G_n(M) / \approx$, there is a representative in the pseudo-diagonal form. Therefore, in the search for optimum commutative group codes, it is sufficient to consider only the set of commutative groups such that their matrices are in the form (\ref{pseudo_diagonal}). Let us denote this set by $B_n$. The cardinality of $B_n$ is clearly finite, since $0 \leqslant b_{ij} \leqslant M$.

However, the set $B_n$ still has equivalent groups and can be reduced. For instance, let $G \in B_n$ be a group of matrices free of $2 \times 2$ reflection blocks, i.e., the elements in $G$ have only $2 \times 2$ rotation matrices as diagonal blocks.

Let 

$$G_{ij}=\left[\begin{smallmatrix} \cos\big(\frac{2\pi b_{ij}}{M}\big) & -\sin\big(\frac{2\pi b_{ij}}{M}\big)\\ \sin\big(\frac{2\pi b_{ij}}{M}\big) & \cos\big(\frac{2\pi b_{ij}}{M}\big) \end{smallmatrix}\right]$$ be the $i$-th block of the $j$-th generator of $G$. The block $G_{ij}$ is a rotation by an angle of $\left(2 \pi b_{ij}/M \right)$. Note that the rotation block corresponding to $M-b_{ij}$ is a conjugate of the block associated with $b_{ij}$: 

$$\left[\begin{smallmatrix} \cos\big(\frac{2\pi (M-b_{ij})}{M}\big) & -\sin\big(\frac{2\pi (M-b_{ij})}{M}\big)\\ \sin\big(\frac{2\pi (M-b_{ij})}{M}\big) & \cos\big(\frac{2\pi (M-b_{ij})}{M}\big) \end{smallmatrix}\right]=\left[\begin{smallmatrix} \cos\big(\frac{2\pi b_{ij}}{M}\big) & \sin\big(\frac{2\pi b_{ij}}{M}\big)\\ -\sin\big(\frac{2\pi b_{ij}}{M}\big) & \cos\big(\frac{2\pi b_{ij}}{M}\big) \end{smallmatrix}\right]=\left[\begin{smallmatrix} 1 & 0  \\ 0 & -1  \end{smallmatrix}\right] G_{ij} \left[\begin{smallmatrix} 1 & 0  \\ 0 & -1  \end{smallmatrix}\right].$$ Therefore, up to conjugacy, we can consider $b_{ij} \leq M/2$ in (\ref{pseudo_diagonal}).

Moreover, the permutation of two consecutive blocks $G_{ij}$ and $G_{(i+1)j}$ (and hence any two rotation blocks) results also in a conjugacy in $\OO_{2k}$. We next consider only the set $B_n$ and also discard equivalent groups, as described above.

In \cite{BigEl03}, Biglieri and Elia present the estimation $\displaystyle \genfrac{(}{)}{0pt}{}{M/2}{n/2}$ for the number of cyclic groups which must be checked in order to find an optimum one. By discarding isometric codes, as stated above, and also considering the \textit{Ádáms' condition} \cite{Adam}, presented next, we can give a lower estimate for the number of cases to be tested in the search for an optimum cyclic group code.

\textbf{Ádáms' condition:} for a fixed $M$ and $a,b \in \Z^k$ we say that $\aa$ and $\bb$ are \linebreak \textit{Ádám's-equivalent} and denote by $\aa \simeq \bb$ iff there exists $\alpha$ invertible in $\Z_M$ \linebreak such that $\aa=\alpha \bb \mod M$. 

A generator matrix of a cyclic group $G \in \OO_{2k}$ can be defined by a vector $\bb=\big(b_1,b_2,\cdots, b_k \big)$ with $0 < b_i \leq M$ and $\gcd(b_1,b_2,\cdots, b_k,M)=1$ to represent the rotation blocks. The Ádám's relation, $\aa \simeq \bb$, implies that two pseudo-diagonal matrices (\ref{pseudo_diagonal}) with parameters defined by $\aa$ and $\bb$ generate the same cyclic group.

Thus, the number of distinct cyclic groups is clearly less than $ \displaystyle \genfrac{(}{)}{0pt}{}{M/2}{n/2}$ and depends on the number of invertible elements in $\Z_M$, which is given by the \textit{Euler phi} function of $M$, $\varphi(M)$. Moreover, as pointed out above, we can restrict our search to vectors $\bb=\big(b_1,b_2,\cdots, b_k \big)$, $0 \leq b_i \leq M/2$. Based on these arguments, we can estimate the number of cyclic group codes, up to symmetry, by $(M/2)^k / \varphi(M)$, which is lower than the number $ \displaystyle \genfrac{(}{)}{0pt}{}{M/2}{n/2}$, given in \cite{BigEl03}. Table \ref{tab_estim} shows a comparison of these values for $k=2$ and several values of $M$. The final column refers to the number of cyclic group codes effectively tested by \textbf{Algorithm 1} (derived in Section 4), which discards additional isometric groups in order to find an optimum code.

\begin{table}[htb]
\begin{center} 
\caption{\label{tab_estim} Different estimations for the number of distinct (non-isometric) order $M$ cyclic group codes in $\R^4$ and cases effectively tested by \textbf{Algorithm 1}.}
$
\begin{array}{|c|c|c|c|}
\hline M  &  \genfrac{(}{)}{0pt}{}{M/2}{n/2} &  \dfrac{M^2}{4\varphi(M)} & \mbox{Algorithm 1} \\ \hline
 32  & 120  & 16  & 14 \\ \hline
 64  & 496  & 32 & 26 \\ \hline
 128 & 2016 & 64 & 50 \\ \hline
 256 & 8128 & 128 & 98 \\ \hline
 512 & 32640 & 256 & 194 \\ \hline
 1024 & 130816 & 512 & 386\\ \hline
\end{array}
$
\end{center}
\end{table}

In what follows, we will focus our attention on the class of commutative group codes, with generator matrices are free of $2 \times 2$ reflection blocks. Moreover, it is sufficient to consider commutative group codes in even dimensions, because, as pointed out in \cite{SIQ}, a commutative group code in odd dimension, $n=2k+1$, is generated by a group $G \in \OO_{2k+1}$ with matrices $g_i \in G$ have a form:

$$	
g_i=[R_1(i),\ldots,R_k(i),\pm 1], \forall  \ \ 1 \leq i \leq M.
$$
This implies that, in a code $\C(M,2k+1)$, the order $M$ must be even, and the code is a union of two $\C(\frac{M}{2},2k)$ contained in parallel hyperplanes. Thus, an optimum commutative group code of order $M$ in $R^{2k+1}$ can be determined starting from the known optimal code in the previous dimension, $\C(\frac{M}{2},2k)$, with initial vector \linebreak $ x_0 = (\delta_1, 0, \dots , \delta_{k}, 0)$. The search for the best initial vector $y_{\theta} = (\cos{\theta} x_0,\sin{\theta})$ is then reduced to a single-parameter optimizing problem.

The next section is devoted to the development of a method for searching for an optimum commutative group code free of reflection blocks in even dimensions. Besides providing additional reduction in the number of cases to be tested, we show how to select and efficiently store a set of cases which allows to finding of an optimal code by solving the correspondent initial vector problems.

\subsection{Describing non-isometric commutative group codes}
\label{sec_desc}

Our approach starts with the connection between commutative group codes and lattices \cite{SIQ}. Speci\-fically, let $\C$ be a commutative group code in $\R^{2k}$, generated by a group $G \in B_n/\approx$, with matrices are free of $2 \times 2$ reflection blocks. We define the associated lattice $\Lambda_G$ by

$$
\Lambda_G := \left\lbrace (b_1, \hdots, b_k) \in \Z^k: [R(b_1), \hdots, R(b_k)] \in G \right\rbrace,
$$
where $R(b)$ denotes the rotation in $\R^2$ by an angle of $2 \pi b/M$ and $[R(b_1), \hdots, R(b_k)]$ denotes a pseudo-diagonal matrix, according to  (\ref{pseudo_diagonal}).

We point out that $\Lambda_G$ contains $M \Z^k:= \left\lbrace M (z_1,z_2,\hdots,z_k), z_i \in \Z \right\rbrace$ as a sub-lattice. Inside the hyperbox $[0,M)^k$ there are exactly $M$ points of $\Lambda_G$, which \linebreak correspond to representatives of the elements of $G$, i.e., 

$$[0,M)^k \supset \left\lbrace (b_{i1},b_{i2},\hdots, b_{ik})\mod M, \ \ i=1,2,\hdots,M \right\rbrace.$$

The lattice $\Lambda_G$ can then be viewed as the translation of these representatives through the lattice $M\Z^k$.

If $x_0=(\delta_1, 0, \dots , \delta_{k}, 0)$ is an initial vector for the code $\C$, we can also define a lattice $\Lambda_G(x_0)$ by

$$
\Lambda_G(x_0) := \left\lbrace \left[\begin{smallmatrix} \frac{2 \pi \delta_1}{M} & & & \\ & \frac{2 \pi \delta_2}{M} & & \\ & & \ddots & \\ & & & \frac{2 \pi \delta_k}{M} \end{smallmatrix}\right] b: b \in \Lambda_G \right\rbrace.
$$

Under these conditions, the code $\C$ is the image $\psi_{x_0}(\Lambda_G(x_0)) \subset S^{2k-1}$, where

\begin{equation}
\label{lambdaGG}
\psi_{x_0}(y)=\displaystyle \left(\delta_{1}\cos\left(\frac{y_{1}}{\delta_{1}}\right),\delta_1\sin\left(\frac{y_{1}}{\delta_{1}}\right),\ldots,\delta_{k}\cos\left(\frac{y_{k}}{\delta_{k}}\right),\delta_k\sin\left(\frac{y_{k}}{\delta_{k}}\right)\right)
\end{equation}
is the standard parametrization of the torus with radii $\delta_i$ \cite{SIQ}.

We say that two lattices $\Lambda_G$ and $\Lambda_H$ are equivalent, and denote by $\Lambda_G \sim \Lambda_H$ iff $\psi(\Lambda_G(x_0))$ and $\psi(\Lambda_H(y_0))$ are isometric codes, for some $x_0$, $y_0$ $\in S^{2k-1}$.

As a consequence of the relation $\sim$, we proceed to use isometry to discard commutative group codes to be checked in the searching for an optimum one. This will be done in terms of basis reduction of the associated lattices, based on results derived in Theorems \ref{FNH} and \ref{Class}.

Theorem \ref{FNH} is closely related to a classical Hermite result. In particular, we have shown that the columns of the resulting matrix $T$ can be ordered by the $\gcd$ (greatest common divisor) of their elements. In Theorem \ref{Class}, we show that it is sufficient to consider generator matrices of lattices in a specific triangular form.

Let $M_k(\Z)$ be the set of $k \times k$ matrices with integer elements. $GL_k(\Z) \subset M_k(\Z)$ is the group of those matrices which are invertible in $M_k(\Z)$, the so called unimodular matrices.

\begin{teo}[Special Hermite Normal Form]
\label{FNH}

Let $B$ be a $k\times k$ matrix with elements in $\Z$. Then there is an upper triangular matrix $T=U\, B\, V$, with $U \in GL_k(\Z)$ and $V$ a permutation matrix. Moreover, $T$ satisfies the following conditions:
\begin{enumerate}
	\item $0 < T(i,i) \leqslant T(i+1,i+1)$, \, $\forall \ \  1 \leqslant i \leqslant k-1$;
	\item $0 \leqslant T ([1:i-1],i) < T(i,i), \, \forall \ \  2 \leqslant i \leqslant k$;
	\item $T(i,i) \leqslant \gcd \left( T([i:j],j) \right), \ \ \forall \ \  1 \leqslant i < j \leqslant k$;
\end{enumerate}
where $T([p:q],r)$ are the elements in the rows $p$ to $q$ of the $r$-th column of $T$.
\end{teo}

\dem

The proof is made by induction on $k$. For $k=1$ it is trivial. Suppose the statement is valid for $n<k$.

Let $V_1$ be a matrix which permutes the columns of $B$, such that the $\gcd$ of the column elements of the matrix $B\,V_1$  are in increasing order.

Let $d_1=\gcd((B\,V_1)_{i,1})$ be the $\gcd$ of the elements in the first column of $B\,V_1$, and $\tilde U_1$ be a unimodular matrix,  such that 

\begin{equation}\label{et1}
\tilde U_1\, B\,V_1 =\left.  \left[\begin{array}{c} d_1 \\ a_2\,d_1\\ \vdots \\ a_k\,d_1 \end{array}\right| \bar{B}_{k,k-1} \right],
\end{equation}
i.e., the product of its first row by the first column of $B\,V_1$ is equal to $d_1$.

Let

\begin{equation}\label{et3}
\hat U_1=\left[\begin{array}{c|c} 1 & \begin{array}{ccc} 0 & \cdots & 0 \end{array}  \\ \hline \begin{array}{c} -a_2 \\ \vdots \\ -a_k \end{array}  & I_{k-1}\end{array}\right], 
\end{equation}
be the matrix which provides the Gaussian elimination in the first column of $B V1$. We thus obtain 

\begin{equation}\label{et2}
\underbrace{\hat U_1\tilde U_1}_{=U_1}\, B\,V_1 =\left.  \left[\begin{array}{c} d_1 \\ 0\\ \vdots \\ 0 \end{array}\right| \tilde B_{k,k-1} \right]. 
\end{equation}

Let $B_1$ be the $(k-1)\times(k-1)$ submatrix of $U_1\, B\,V_1$, obtained by removing the first row and first column. By the induction hypothesis there exists a unimodular matrix $\tilde U$ and a permutation matrix $\tilde V$ such that $\tilde T=\tilde U\, \tilde B_1 \, \tilde V$.

Then, 
{\footnotesize

\begin{eqnarray}\label{et4}
\left[ \begin{array}{cc} 1 & 0 \\  0 & \tilde U \end{array} \right]\, U_1 \, B \, V_1 \, \left[ \begin{array}{cc} 1 & 0 \\  0 & \tilde V \end{array} \right] & = & 
\left[ \begin{array}{cc} 1 & 0 \\  0 & \tilde U \end{array} \right]\, \left[\begin{array}{cc} d_1 & e  \\ 0 & B_1\end{array} \right] \, \left[ \begin{array}{cc} 1 & 0 \\  0 & \tilde V \end{array} \right] \nonumber \\
& = &  \left[ \begin{array}{cc} d_1 & e \\  0 & \tilde U \, B_1 \end{array} \right] \, \left[ \begin{array}{cc} 1 & 0 \\  0 & \tilde V \end{array} \right] \nonumber \\
& = &  \left[ \begin{array}{cc} d_1 & e\, \tilde V \\  0 & \tilde U \, B_1\,\tilde V  \end{array} \right] \nonumber \\
& = &  \left[ \begin{array}{cc} d_1 & e\, \tilde V \\  0 & \tilde T \end{array} \right] = T.
\end{eqnarray}
}

If $T(1, j) <0$ or $T(1, j) > T(j, j)$  for some $j>1$, we can apply the elementary operation $\ell_1 \leftarrow \ell_1 - \left \lfloor \dfrac{T_{1, j}}{T_{j, j}} \right \rfloor  \ell_j$, or equivalently left-multiply $T$ by a unimodular matrix $\bar U_j$, to conclude the proof.

\fim

In contrast to the standard Hermite normal form \cite{Cohen}, here the unimodular matrix $U$ is operating on the left side of $B$. In other words, if the rows of $B$ contain the generator vectors of a $k$-dimensional lattice, then matrix $U$ represents a change of basis in this lattice. Moreover, the permutation matrix $V$, which does not appear in the standard Hermite normal form, allows us to sort the columns of $T$ by their greatest common divisor, which will be useful in order to discard isometric codes. We remark that the matrix $V$, operating on the right side of $B$, represents an isometry by coordinate permutation. Thus the lattices generated by $T$ and $B$ can be different, but they are isometric.

\begin{teo}
\label{Class}
Every commutative group code $\C \subset S^{2k-1}$, generated by a group $G \in \OO_{2k}$ free of $2 \times 2$ reflection blocks is isometric to a code obtained as image by $\psi$ of a lattice $\Lambda_G(x_0)$. Moreover the associated lattice $\Lambda_G$ has a generator matrix $T$ satisfying the following conditions:
\begin{enumerate}
	\item $T$ is upper triangular according to Theorem \ref{FNH};
	\item $\det(T) = M^{k-1}$;
	\item There is a matrix $W$, with integer elements satisfying $W\,T=M\,I_k$, where $I_k$ is the $k \times k$ identity matrix;
	\item The elements of the diagonal of $T$ satisfy $\displaystyle T(i,i)=\frac{M}{a_i}$  where $a_i$ is a divisor of $M$ and $(a_{i})^{i} \cdot (a_{i+1} \cdots a_k) \leqslant M$, $\forall i=1, \hdots, k$.
	\end{enumerate}
\end{teo}

\dem

\textit{1 - } Let $B$ be a generator matrix of the lattice $\Lambda_T$. By Theorem \ref{FNH}, there exists an upper triangular matrix $T$ such that $T=U\,B\,V$. Since the matrix $U$ is unimodular, it defines a change of basis in the lattice generated by $B$, while $V$ is an isometry  by coordinate permutation. Both operations are isometries in lattices, thus, matrices $B$ and $T=U\,B\,V$ define lattices which are equivalent and which, therefore, generate isometric commutative group codes.

\textit{2 - } The lattice $\Lambda_G$ contains the sublattice $M \Z^k$ and the cardinality of the quotient $\displaystyle \frac{\Lambda_G}{M \Z^k}$ must be equal to $M$, the number of points in the code. Therefore, since $\det(M \,I_k)=M^k$ we conclude that $\det(T) = M^{k-1}$.

\textit{3 - } The system $x \,T= M\,e_i$ must have a solution in $\Z^k$ for all $1 \leqslant i \leqslant k$, where $e_i$ is the $i$-th column of $I_k$. Let $W$ be the matrix with rows containing these solutions; then $W \, T = M I_k$. \footnote{Note that condition $3$ is equivalent to saying that $(M \Z^k)$ is a sublattice of $\Lambda_G$.}

\textit{4 - } The number $M$ must be a multiple of the elements in the diagonal of $T$ (from item \textit{2}) moreover, from Theorem \ref{FNH}, we know that

$$T(i,i) \leqslant T(i+1,i+1) \, \mbox{ and then } \frac{M}{a_i} \leqslant \frac{M}{a_{i+1}} \mbox{ which implies that } a_{i+i} \leqslant a_{i}.$$
From 

$$\displaystyle \det(T)= \frac{M}{a_{1}} \, \frac{M}{a_{2}}\, \hdots \frac{M}{a_{k}} = M^{k-1},$$ we get

$$
(a_1 \, a_2 \, \cdots \, a_k) = M \Rightarrow (a_{i})^{i} \cdot (a_{i+1} \cdots a_k) \leqslant M.
$$
\fim

Not all upper triangular integer matrices $T$ satisfy the conditions of Theorem \ref{Class}. For example, for $M=12$ and $k=3$, the matrix

$$T=\left[ \begin{array}{ccc}
2 & 3 & 0 \\ 
0 & 6 & 6 \\ 
0 & 0 & 12
\end{array}\right] , $$
satisfies the hypothesis of Theorem \ref{FNH} and  $\det(T)=12^2$ but, in order to obtain $W\,T=12\,I_3$, we must have:

$$W=\left[ \begin{array}{ccc}
6 & -3 & 3/2 \\ 
0 & 2 & -1 \\ 
0 & 0 & 1
\end{array}\right].$$
However, in this case, $W$ has non-integer elements.

In order to characterize a commutative group code as an image of a quotient of lattices, it is also important to determine a set of generators of the correspondent group and its class of isomorphism. In the Theorem \ref{gera} we deal with this problem.

\begin{teo}[\cite{Cohen}, p 76]
\label{FNS}
Let $A$ be a non-singular $k\times k$ matrix with coefficients in $\Z$. There is then a unique diagonal matrix $D=(d_{i,j}),$ with $d_{i+1,i+1}|d_{i,i},$ such that $D=V\, A\, U$ with $U$ and $V$ in $GL_k(\Z)$.
\end{teo}

This matrix is called the \textit{Smith normal form} (SNF) of A. 
 
\begin{teo}[\cite{Cohen}, p 76]
\label{EDT2}
Let $L$ be a $\Z$-submodule of a free module $L^\prime$ and of the same rank. Then there are positive integers $d_1, \dots , d_k$ satisfying the following conditions:
\begin{enumerate}
\item For every $i$ such that $1\leqslant i < k$ we have $d_{i+1}|d_i.$ 
\item As $\Z$-modules, we have the isomorphism

$$L^\prime/L \simeq \bigoplus_{1\leqslant i \leqslant k} (\Z/d_i\, \Z) = \bigoplus_{1\leqslant i \leqslant k} (\Z_{d_i})$$
and in particular $[L^\prime:L]=d_1 \cdots d_k$ and $d_1$ is the exponent of $\frac{L^\prime}{L}$.
\item There is a $\Z$-basis $\{\vetn{v}{k}\}$ of $L^\prime$ such that $\{d_1 \vv v_1, \dots , d_k \vv v_k\}$ is a $\Z$-basis of $L$.
\end{enumerate}
\end{teo}

\begin{teo}
\label{gera}
For a commutative group code $\C$, let $T$ be a generator matrix of the lattice $\Lambda_G$, according Theorem \ref{Class} and $W=M T^{-1}$. The set of generators of the correspondent group and its class of isomorphism are then obtained from the SNF of $W$.
\end{teo}

\dem

Let $D=V\, W\, U$ be the SNF of $W$. We know that 

$$W\,T = M\,I_k \Rightarrow V^{-1}\,D\,U^{-1} = M\,I_k \Rightarrow D\,U^{-1}\,T=V\,M\,I_k.$$ Since the matrices $U^{-1}$ e $V$ are unimodular, their product on the left of the generator matrices $T$ and $M \, I_k$ define a change of basis in the  lattice generated by $T$ and its sublattice $M \Z^k$. The classification and generators of the group are derived from Theorem \ref{EDT2}. In this case, $G$ is isomorphic to a group $ \Z_{d_1} \otimes \hdots \otimes \Z_{d_k}$ and the rows of $U^{-1}\,T$ give the elements $b_{ij}$ which form a set of generators, according to (\ref{pseudo_diagonal}).

\fim

As a consequence of these results, we derive a two-step algorithm which searches for an optimum commutative group code $\C$ of order $M$ in an even dimension. The first step consists of storing all matrices $T$ according to theorem \ref{Class} and the use of Ádám's relation to discard isometric groups. For each one of these matrices $T$ we then establish a linear programming problem (Section 3) to determine the initial vector $x_0$ which maximizes the minimum distance of the group code $\psi_{x_0}\Lambda_G(x_0)$ (\ref{lambdaGG}). For the optimum case, theorem 4.5 is applied to obtain the generators and the class of isomorphism of the commutative group. The algorithm is summarized as a pseudo code in \textbf{Algorithm 1}.

\begin{algorithm}
\SetKwData{Left}{left}\SetKwData{This}{this}\SetKwData{Up}{up}
\SetKwFunction{Union}{Union}\SetKwFunction{FindCompress}{FindCompress}
\SetKwInOut{Input}{input}\SetKwInOut{Output}{output}
\Input{The number of points $M$ and the dimension $n=2k$;}
\Output{An optimum commutative group code $\C(M,n)$, its set of generators, optimum initial vector, its isomorphism class and minimum distance.}
\BlankLine
\Begin{
$dist \leftarrow 0$;\\
$div \leftarrow \left\{ a_1,a_2, \hdots, a_w \right\}$, the set of divisors of $M$;\\
$A \leftarrow \left[ diag_1 |, diag_2|, ..., diag_j| \right]$, a matrix with columns contains all the possible diagonals for $T$, according Theorem  \ref{Class}, i.e.,  $diag_i = \big( \frac{M}{a_{i,1}}, \frac{M}{a_{i,2}}, \hdots, \frac{M}{a_{i,k}} \big)^t, \ \ \mbox{ where } a_{i,k} \in div$, $a_{i,k} \geq a_{i,k+1}$ and $\prod_{q=1}^k a_{i,q} = M$;\\
\ForEach{$diag_i \in A$}{
\textbf{Step 1:} Construct all matrices $T$, according Theorem \ref{Class} and use Ádám's relation to discard isometric groups;\\
\ForEach{matrix $T_i\xi$ constructed in step 1}{\label{forins}
\textbf{Step 2:} Solve the initial vector problem and get the minimum distance $dist_{i\xi}$ and the initial vector $x_{0_{i\xi}}$;\\
\If{$dist_{i\xi} > dist $}{
			   					   		$dist \leftarrow dist_{i\xi}$;\\
			   					   		$x_0  \leftarrow x_{0_{i\xi}}$;\\
			   					   		$T \leftarrow  T_{i\xi}$}
			   					   		}
			   					   		}
Apply Theorem \ref{gera} in $T$ and get the generator of the group $G \in \OO_n$ and the correspondent isomorphism class; 

Output $G$, $x_0$, $dist$ and the isomorphism class.
			   					   		}
\caption{Optimum commutative group code}\label{algo_disjdecomp}
\end{algorithm}
Let us illustrate this method in detail for $M=128$ and $n=4$.

Let $div = \left\{ 1, 2, 4, 8, 16, 32, 64, 128\right\}$ be the set of divisors of $128$. From Theorem \ref{Class}, we know that the matrix $T$, related to a code $\C(128,4)$, has the form

$$
T =
\left[ 
\begin{array}{cc}
d_1 & w \\ 
0 & d_2
\end{array} 
\vspace{0.4cm}
\right], \mbox{with } d_i = \frac{M}{a_i}, \\  {a_i} \in div.
$$ 
Moreover $(d_2)^2 < 128$, i.e., $ d_2 \in \left\lbrace 1,2,4,8 \right\rbrace $. We can then store all the possible diagonal of $T$ as columns of a matrix $A$:

$$A = 
\left[
\begin{array}{cccc}
1&2&4&8\\
128&64&32&16
\end{array} 
\right].
$$
For each column of $A$, the set of values $w$ in $T$ can then be determined, as established in item \textit{3.} of Theorem \ref{FNH}, by considering 

$$a_{1_i} \leqslant \gcd(w,a_{2_i}),$$  In this example, we have
 
$
T_1 =
\left( 
\begin{array}{cc}
1 & w_{1 \xi} \\ 
0 & 128
\end{array} 
\vspace{0.4cm}
\right), 
\mbox{ where } w_{1\xi} \in \left\lbrace 0,1, \cdots, 64  \right\rbrace ;
$

$
T_2 =
\left( 
\begin{array}{cc}
2 & w_{2\xi} \\ 
0 & 64
\end{array} 
\vspace{0.4cm}
\right), 
\mbox{ where } w_{2\xi} \in \left\lbrace 0,2,4,6, \cdots, 32  \right\rbrace ;
$

$
T_3 =
\left( 
\begin{array}{cc}
4 & w_{3\xi} \\ 
0 & 32
\end{array} 
\vspace{0.4cm}
\right), 
\mbox{ where } w_{3\xi} \in \left\lbrace 0,4,8,12,16  \right\rbrace ;
$

$
T_4 =
\left( 
\begin{array}{cc}
8 & w_{4\xi} \\ 
0 & 16
\end{array} 
\vspace{0.4cm}
\right), 
\mbox{ where } w_{4\xi} \in \left\lbrace 0,8  \right\rbrace .
$

This amounts to $89$ cases to be tested. However some of these lattices are equivalent. For example, the lattice generated by a matrix $T_1$ which has the first row equal to $(1,w_{1 \xi})$ is equivalent to a lattice generated by a matrix $T_1$ which has the first row equal to $(1,w_{1 \xi}^{-1})$, here $w_{1 \xi}^{-1}$ represents the inverse of $w_{1 \xi}$ in $\Z_{128}$. If $w_{1\xi}^{-1} < w_{1\xi}$, we can therefore discard the correspondent matrix in set $T_1$. This situation occurs for 

$$w_{1 \xi}=\left\lbrace 17,27,33,35,39,41,43,45,49,51,53,55,57,59,61 \right\rbrace.$$ 

Similarly, the lattice generated by a matrix $T_2$, which has the first row equal to $(2, 2b)$ is equivalent to a lattice generated by a matrix $T_2$ which has the first row equal to $(2,2b^{-1})$. Thus, in this set we can discard the cases where $w_{2\xi}=\left\lbrace 18,26 \right\rbrace$. Therefore, in order to find an optimum code $\C(128,4)$ it is sufficient to check $72$ codes.

In the implementation of \textbf{Algorithm 1}, these equivalent cases can be discarded during \textit{Step 1} and the solution of the initial vector problem, consequently, implemented just for the relevant cases. Only the matrix which determines the largest minimum distance must be saved.

In this example, the optimum code is associated to the matrix
$
T_{1,12} =
\left[ 
\begin{array}{cc}
1 & 11 \\ 
0 & 128
\end{array} 
\vspace{0.4cm}
\right] .
$

The correspondent group $G\in \OO_n$ is then obtained using the SNF of \linebreak $W = M \left(T_{1,12}\right)^{-1}$.

In this case, the best $\C(128,4)$ is a cyclic group code with the following generator matrix:

$$
{\footnotesize
G_{(1,11,128)} = \left[
\begin{array}{cccc}
 \cos{\left(\dfrac{1*2 \pi}{128}\right)} & \sin{\left(\dfrac{1*2 \pi}{128}\right)} & 0 & 0 \\
-\sin{\left(\dfrac{1*2 \pi}{128}\right)} & \cos{\left(\dfrac{1*2 \pi}{128}\right)} & 0 & 0 \\
 0 & 0 & \cos{\left(\dfrac{11* 2 \pi}{128}\right)} & \sin{\left(\dfrac{11* 2 \pi}{128}\right)} \\
 0 & 0 & -\sin{\left(\dfrac{11 *2 \pi}{128}\right)} & \cos{\left(\dfrac{11 *2 \pi}{128}\right)}
\end{array}
\right]}.
$$  

The minimum distance in this code is $d = 0.406179$ for the best initial vector \linebreak $x_0 = ( 0.65098, 0, 0.759095, 0)^t$.

In dimension $4$, the number of commutative group codes tested by Algorithm 1 is not much larger than the number of cyclic group codes tested (Table \ref{tab_estim}). For $M$ equals to $32$ (respectively, $64, 128, 256, 512, 1024$), Algorithm 1 checks $21$ (respectively, $38, 72, 141, 273, 542$) commutative group codes in order to find an optimum one.

\begin{figure}[h!]
\centering
		\includegraphics[width=12cm]{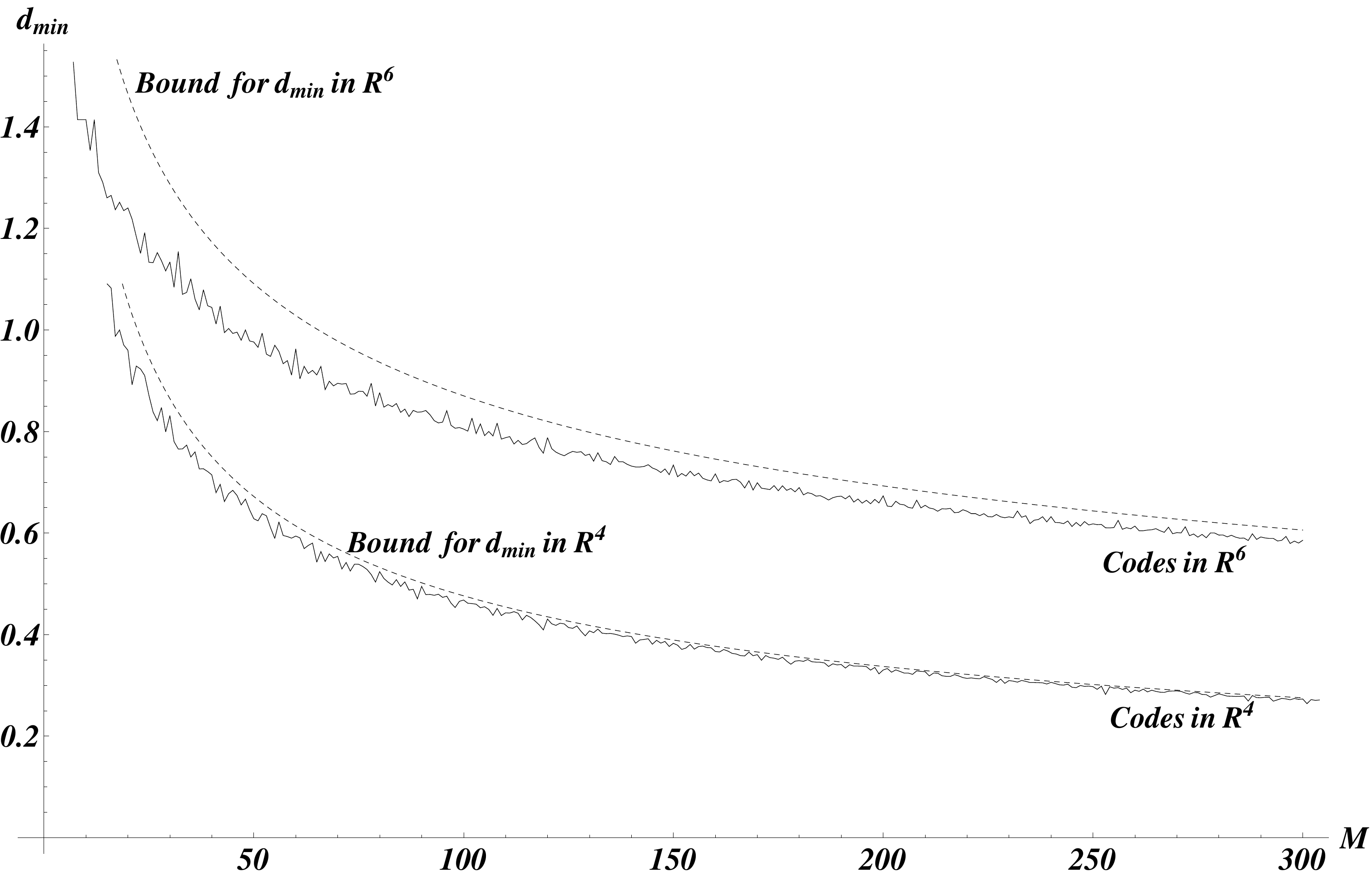}
	\caption{Comparison between the distance of optimal codes found using Algorithm 1 and upper bound \cite{SIQ}: the gap decreases when $M$ grows.}
		\label{fig.graf}
\end{figure}

Using this method we have found optimum codes for various values of $M$ in different dimensions and we have present some of them in $\R^4$ and $\R^6$ in Tables \ref{tabela2} and \ref{tabela3}. In both cases, it can be seen that, when the number of points $M$ increases, the gap between the minimal distance of the codes and the upper bound \cite{SIQ} decreases. This fact is also illustrated in Figure \ref{fig.graf}.

\begin{table}[htb]
\begin{center}
\footnotesize
\caption{\label{tabela2} Some optimum commutative group codes of order $M$ in $R^4$.}
\begin{tabular}{|c|c|c|c|c|c|c|}
\hline  $M$    &  $d_{min}$   &  $\delta_1$  &  $\delta_2$  &    Group                     & Gen.  $(b_{ij})$             &  Bound \\
\hline  10   &  1.224  &  0.707  &  0.707  &   $\Z_{10}$                          &(1 3)              &   1.474 \\
\hline  20   &  0.959  &  0.678  &  0.734  &   $\Z_{20}$                          &(3 4)              &   1.054\\
\hline  30   &  0.831  &  0.707  &  0.707  &   $\Z_{30}$                          &(3,5)              &  0.864 \\
\hline  40   &  0.714  &  0.607  &  0.794  &   $\Z_{40}$                          &(4 5)        	    &  0.750 \\
\hline  50   &  0.628  &  0.707  &  0.706  &   $\Z_{50}$                          &(7 2)              &  0.672 \\
\hline 100   &  0.468  &  0.757  &  0.653  &   $\Z_5 \, \oplus \, \Z_{20}$       &(0 20), (5 10)     &  0.476 \\
\hline 200   &  0.330  &  0.750  &  0.660  &   $\Z_{200}$                         &(93 1)             &  0.337 \\
\hline 300   &  0.273  &  0.656  &  0.754  &   $\Z_5 \, \oplus \, \Z_{60}$      &(60 120), (10 15)   &  0.275  \\
\hline 400   &  0.237  &  0.686  &  0.727  &   $\Z_{400}$                         &(189 1)            &  0.238 \\
\hline 500   &  0.211  &  0.674  &  0.738  &   $\Z_{500}$                         &(13 20)            &  0.213 \\
\hline 600   &  0.193  &  0.676  &  0.736  &   $\Z_{600}$                         &(191 198)          &  0.194 \\
\hline 700   &  0.180  &  0.718  &  0.695  &   $\Z_{700}$                         &(14 25)            &  0.180 \\
\hline 800   &  0.168  &  0.670  &  0.742  &   $\Z_{800}$                         &(16 25)            &  0.168 \\
\hline 900   &  0.158  &  0.704  &  0.709  &   $\Z_{900}$                         &(197 2)            &  0.159 \\
\hline 1000  &  0.149  &  0.716  &  0.697  &   $\Z_{1000}$                        &(33 4)             &  0.150 \\
\hline 
\end{tabular}
\end{center}
\end{table}

\begin{table}[htb]
\begin{center}
\caption{\label{tabela3} Some optimum commutative group codes of order $M$ in $R^6$.}
\scriptsize
\begin{tabular}{|c|c|c|c|c|c|c|c|}
\hline  $M$    &  $ d_{min}$   &  $\delta_1$  &  $\delta_2$ & $\delta_3$ &    Group                       & Gen $(b_{ij})$       &  Bound    \\
\hline  10     &  1.414     &  0.632    &  0.632   &  0.447   &  $\Z_{10}$                         & (3,1,5)                  &  1.820 \\
\hline  20     &  1.240     &  0.554    &  0.620   &  0.554   &  $\Z_{20}$                         & (2,5,6)                  &  1.465 \\
\hline  30     &  1.133     &  0.534    &  0.654   &  0.534   &  $\Z_{30}$                         & (3,5, 9)                 &  1.287 \\
\hline  40     &  1.044     &  0.603    &  0.522   &  0.603   &  $\Z_{2} \, \oplus \, \Z_{20} $   & (20,0,20), (32,10,4)     &  1.173 \\
\hline  50     &  0.976     &  0.604    &  0.506   &  0.615   &  $\Z_{50}$                         & (7,6, 34)                &  1.091 \\
\hline 100     &  0.804     &  0.515    &  0.684   &  0.515  &  $\Z_{10}  \, \oplus  \, \Z_{10}$ & (50, 10, 0),  (30, 0, 10) &  0.870 \\
\hline 200     &  0.673     &  0.555    &  0.619   &  0.555  &  $\Z_{200}$                        & (28, 25, 4)               &  0.692 \\
\hline 300     &  0.585     &  0.585    &  0.498   &  0.639  &  $\Z_5 \, \oplus  \, Z_{60}$      & (0, 0, 60), (25, 30, 30)  &  0.605 \\
\hline 400     &  0.540     &  0.562    &  0.605   &  0.562  &  $\ Z_{20} \, \oplus  \, Z_{20}$  & (300, 40, 0), (60, 0, 20) &  0.550 \\
\hline 500     &  0.504     &  0.577    &  0.577   &  0.577   &  $\ Z_5 \, \oplus  \, Z_{10}, \, \otimes  \,Z10 $ & (100, 0, 0), (50, 50, 0), (50, 0, 50)  &  0.511\\
\hline 600     &  0.472     &  0.549    &  0.630   &  0.549  &  $\ Z_{2} \, \oplus  \, Z_{300}$  & (300, 0, 300),  (384, 50, 12) &  0.481\\
\hline 700     &  0.445     &  0.531    &  0.612   &  0.585  &  $\ Z_{700}$                       & (457, 664, 298)          &  0.457\\
\hline 800     &  0.427     &  0.617    &  0.486   &  0.617  &  $\Z_{20} \, \oplus  \, \Z_{40}$  &(80,0,40),(20,80,60)      &   0.437 \\
\hline 900     &  0.413     &  0.592    &  0.591   &  0.547  &  $\Z_3  \, \oplus  \, Z_{300}$    &(0,300,0),(759,36,3)      &  0.420\\
\hline 1000    &  0.397     &  0.560    &  0.632   &  0.535  &  $\Z_{1000}$                       &(319,694,45)              &  0.406\\
\hline
\end{tabular}
\end{center}
\end{table}

Although some other group codes, as permutations codes, can outperform commutative group codes for some parameters \cite{zino}, they are very special for some applications as transmition over symmetric channels \cite{Como09}. Besides, they may provide homogeneous spherical codes for any number of codewords and can be used for designing high density spherical codes on flat torus layers \cite{isit2009}.

\section{Conclusions}
A two-step method for finding an optimum $n$-dimensional commutative group code of order $M$ is presented. The approach explores the structure of lattices associated with these codes in even dimensions and allows a significant reduction in the number of non-isometric cases to be analyzed. For each of these cases, a linear programming problem is solved to find the initial vector which maximizes the minimum distance in the code. The method introduced here can also be used to design more general spherical codes, such as the so called quasi-commutative group codes, which are constructed on layers of flat tori \cite{isit2009}.

\end{document}